\documentclass[12pt]{article}
\usepackage{amsmath}
\numberwithin{equation}{section}
\usepackage{amssymb}
\usepackage{graphicx,psfrag,epsf}
\usepackage{natbib,hyperref}

\setcitestyle{aysep={}}

\usepackage[utf8]{inputenc}
\usepackage[english]{babel}
\usepackage{url} 

\usepackage{setspace}
\usepackage[format=plain, labelfont=bf, singlelinecheck=off]{caption} 
\usepackage{booktabs}
\usepackage{array}
\usepackage{multirow}
\usepackage{xcolor}

\usepackage[toc,page]{appendix}

\newcommand{\blind}{1}

\usepackage[
top    = 1in,
bottom = 1.2in,
left   = 1in,
right  = 1in]{geometry}

\usepackage{enumitem}
\usepackage{bbm}
\usepackage{verbatim}
\usepackage{xcolor}

\newcommand{\s}{\mathbf{s}}

\newcommand{\A}{\mathbf{A}}

\newcommand{\D}{\mathbf{D}}
\newcommand{\V}{\mathbf{V}}

\renewcommand{\H}{\mathbf{H}}
\newcommand{\X}{\mathbf{X}}
\newcommand{\Z}{\mathbf{Z}}
\newcommand{\Y}{\mathbf{Y}}
\newcommand{\Q}{\mathbf{Q}}
\renewcommand{\S}{\mathbf{S}}
\newcommand{\K}{\mathbf{K}}

\newcommand{\boldSigma}{\boldsymbol \Sigma}
\newcommand{\boldxi}{\boldsymbol \xi}
\newcommand{\boldeta}{\boldsymbol \eta}

\newcommand{\boldmu}{\boldsymbol \mu}
\newcommand{\boldbeta}{\boldsymbol \beta}
\newcommand{\boldtheta}{\boldsymbol \theta}
\newcommand{\boldDelta}{\boldsymbol \Delta}

\expandafter\def\expandafter\normalsize\expandafter{%
    \normalsize
    \setlength\abovedisplayskip{6pt}
    \setlength\belowdisplayskip{6pt}
    \setlength\abovedisplayshortskip{6pt}
    \setlength\belowdisplayshortskip{6pt}
}

\newcommand{\bfbeta}{\boldsymbol \beta}
\newcommand{\bfepsilon}{\boldsymbol \epsilon}
\newcommand{\bfeta}{\boldsymbol \eta}
\newcommand{\bfxi}{\boldsymbol \xi}
\newcommand{\bfSigma}{\boldsymbol \Sigma}
\newcommand{\bfDelta}{\boldsymbol \Delta}
\newcommand{\bftheta}{\boldsymbol \theta}
\newcommand{\bfmu}{\boldsymbol \mu}

\newcommand{\bZ}{\mathbf{Z}}
\newcommand{\bS}{\mathbf{S}}
\newcommand{\bA}{\mathbf{A}}
\newcommand{\bC}{\mathbf{C}}
\newcommand{\bD}{\mathbf{D}}
\newcommand{\bI}{\mathbf{I}}
\newcommand{\bH}{\mathbf{H}}
\newcommand{\bK}{\mathbf{K}}
\newcommand{\bQ}{\mathbf{Q}}
\newcommand{\bT}{\mathbf{T}}
\newcommand{\bX}{\mathbf{X}}
\newcommand{\bfzero}{\mathbf{0}}

\usepackage{amsthm}

\newtheorem{proposition}{\textsc{Proposition}}

\begin{document}

\def\spacingset#1{\renewcommand{\baselinestretch}%
{#1}\small\normalsize} \spacingset{1}


\if1\blind
{
  \title{\bf A Fused Gaussian Process Model for Very Large Spatial Data\thanks{Pulong Ma is Postdoctoral Fellow, The Statistical and Applied Mathematical Sciences Institute, 79 T.W. Alexander Drive, P.O. Box 110207, Durham, NC 27709 (pulong.ma@duke.edu); and Emily L. Kang is Associate Professor, Department of Mathematical Sciences, University of Cincinnati, 2815 Commons Way, Cincinnati, OH 45221-0025 (kangel@ucmail.uc.edu).
This work was supported in part by an allocation of computing time from the Ohio Supercomputer Center. This research was part of Ma’s Ph.D. dissertation supported by the Charles Phelps Taft Dissertation Fellowship at the University of Cincinnati. Ma was also partially supported by the National Science Foundation under Grant DMS-1638521 to the Statistical and Applied Mathematical Sciences Institute. Kang was supported by the Simons Foundation's Collaboration Award (\#317298) and the Taft Research Center at the University of Cincinnati. Any opinions, findings, and conclusions or recommendations expressed in this material do not necessarily reflect the views of the National Science Foundation. We thank the Associate Editor and anonymous referees for the criticism that improved the paper.}
}
  \author{Pulong Ma
\hspace{.2cm}\\
    Statistical and Applied Mathematical Sciences Institute and 
    Duke University\\
    and \\
    Emily L. Kang \\
    Division of Statistics and Data Science,\\ Department of Mathematical Sciences, University of Cincinnati}
    \date{}
  \maketitle
} \fi

\if0\blind
{
  \bigskip
  \bigskip
  \bigskip
  \begin{center}
    {\LARGE\bf A Fused Gaussian Process Model for Very Large Spatial Data}
\end{center}
  \medskip
} \fi

\bigskip
\begin{abstract}
With the development of new remote sensing technology, large or even massive spatial datasets covering the globe become available. Statistical analysis of such data is challenging. This article proposes a semiparametric approach to model large or massive spatial datasets. In particular, a Gaussian process with additive components is proposed, with its covariance structure consisting of two components: one component is flexible without assuming a specific parametric covariance function but is able to achieve dimension reduction; the other is parametric and simultaneously induces sparsity. The inference algorithm for parameter estimation and spatial prediction is devised. The resulting spatial prediction methodology that we call fused Gaussian process (FGP), is applied to simulated data and a massive satellite dataset. The results demonstrate the computational and inferential benefits of FGP over competing methods and show that FGP is robust against model misspecification and captures spatial nonstationarity. The supplemental materials are available online.\end{abstract}

\noindent%
 \vfill

\newpage
\spacingset{1.45} 
\section{Introduction}
\label{sec:intro}
\doublespacing

With the advances of remote sensing technologies, massive scientific data can be collected over (a proportion of) the globe. Such spatially correlated datasets allow researchers to investigate various issues in environmental and atmospheric sciences. Classic statistical methods such as kriging have been widely used to model spatial data \citep{Cressie1993, Cressie2011, Banerjee2014}. However, with large or massive data,  direct implementation of these statistical methods becomes computationally prohibitive, since solving the kriging equations involves the Cholesky factorization of an $n\times n$ covariance matrix for data of size $n$, which requires computational cost $O(n^3)$ and memory cost $O(n^2)$ in general. 

To tackle these issues, many recent developments in spatial statistics have focused on modeling large or massive spatial datasets. Most of them assume a specific form for the spatial covariance function known up to several parameters, e.g., the Mat\'ern family, and then use different approaches to represent or approximate the target function, the resulting covariance or precision matrix, or the likelihood function. Methods in this paradigm include but are not limited to approximate likelihood \citep[e.g.,][]{Stein2004, Caragea2006}, covariance tapering \citep{Furrer2006, Kaufman2008}, predictive process \citep{Banerjee2008} and its variants \cite{Sang2012}, composite likelihood \citep{Lindsay1988, Eidsvik2014}, nearest-neighbor Gaussian process \citep{Datta2015}, Gaussian Markov random field representation \citep{Lindgren2011},  a multi-resolution approximation for Gaussian process \citep{Katzfuss2017}, and spectral methods \citep{Guinness2016}. 

While the richness and flexibility of the methods mentioned above are indisputable, their implementation and performance generally rely on the assumption of a particular parametric form for the spatial covariance function. One of the main difficulties in using these methods to analyze massive data observed on a very large spatial domain such as the globe is to choose a specific covariance function that can represent various spatial structures in data, since misspecification of the spatial covariance function can have a large impact on inferential efficacy. Therefore, a family of covariance functions that are more flexible and robust is more desirable. Some extensions of the aforementioned methods to construct complex spatial processes have been proposed as well. For instance, by specifying a nonstationary Mat\'ern covariance function in \cite{Paciorek2006}, random spatial basis functions are obtained through a reversible-jump Markov chain Monte Carlo algorithm in \cite{Katzfuss2013}, which can increase the flexibility of the resulting spatial process constructed from the predictive process. Another flexible and nonstationary covariance function is developed by adaptively partitioning the spatial domain through a Bayesian treed Gaussian process in \cite{Konomi2014} and \cite{Konomi_NSNNGP2019}. However, these extensions in general result in much more complicated algorithms, and the resulting computational cost can be too demanding for massive spatial data sets.

A second avenue of recent research has been focused on semiparametric models for analyzing large or massive spatial data sets. Methods in this paradigm represent a spatial process as a linear combination of multiresolutional basis functions and random coefficients \citep[e.g.,][]{Cressie2006, Cressie2008, Chu2014}. Specifically, a flexible family of nonstationary spatial covariance functions is developed in \cite{Cressie2008} based on a pre-specified multiresolutional and compactly supported basis functions and a general low-rank covariance matrix. This model and resluting method is called fixed rank kriging (FRK) in \cite{Cressie2008}. Based on a novel local Karhunen-Lo\`eve expansion for an underlying spatial process, another type of flexible and unspecified spatial covariance functions is proposed in \cite{Chu2014}, where the consistency conditions for parameter estimators in the covariance function are established to estimate a general covariance matrix. However, these semiparametric models have their own limitations. The FRK model incorporates a small number of basis functions to model a spatial process. The resulting low-rank covariance function allows fast computations but may also incur sacrifices in capturing the spatial structures at various scales presented in data \citep[e.g.,][]{Stein2014}. \cite{Chu2014} partition the spatial domain into identically shaped subdomains and assume independence among different subdomains, but the resulting loss of accuracy needs to be recovered by other approaches in practice such as tapering approximation. Instead of using a low-rank structure, \cite{Nychka2015} propose a model called Lattice Krig in which they use a large number basis functions from multiple resolutions. It is then assumed that random effects corresponding to different resolutions are independent, while random effects from the same resolution follow a Gaussian Markov random field (GMRF) using the result in \cite{Lindgren2011} that this particular GMRF model is explicitly connected to the Mat\'ern covariance family. \cite{Nychka2015} focus primarily on presenting how well Lattice Krig approximates a stationary covariance function but without simulation results to illustrate its predictive performance when the true covariance is nonstationary or when the parametric form of a covariance function is unknown a priori.

Motivated by the full-scale approximation in \cite{Sang2012} that the covariance matrix is written as a sum of a low-rank matrix and a tapered covariance matrix, our method also represents the covariance matrix as a sum of two components. Different from  \cite{Sang2012}, we do not need to have a prespecified form for the covariance function of a spatial process. Our work combines the advantages of methods from both parametric and semiparametric paradigms. In particular, a spatial process is constructed with additive components, which have two different types of basis representations based on a set of \textit{inducing variables}. The first component is characterized through a semiparametric representation with a relatively small number of basis functions and a general unspecified form of covariance matrix in the same way as in \cite{Cressie2008}. This low-rank component is able to model nonstationary covariance structures. The second component in the covariance structure is defined through a Gaussian graphical model (GGM), also called Gaussian Markov random field in spatial statistics. We will demonstrate that several state-of-the-art methods for parametric covariance approximation can be viewed as special cases of the GGM under various assumptions. The resulting method is called the fused Gaussian process (FGP) as it blends two components in its covariance function that induces a low-rank matrix and a sparse precision matrix. Unlike the full-scale approximation in \cite{Sang2012} that two components of covariance structures are used together to approximate a target covariance function specified a priori, the low-rank and GGM components in the FGP do not require pre-specification of a particular parametric covariance.  By taking advantage of the properties from both components, we develop a computationally efficient algorithm for parameter estimation and spatial prediction. The superior performance of the proposed approach is demonstrated through simulation studies and a massive satellite dataset. 

The remainder of the paper is organized as follows. Section \ref{sec: model} presents the semiparametric statistical model and discusses about relevant model specification. In Section \ref{sec: inference}, likelihood-based inference including parameter estimation and spatial prediction is devised for the FGP.  
Section \ref{sec: numerical examples} uses simulation examples to demonstrate the robustness of predictive performance and nonstationary performance of the FGP. In Section~\ref{sec: application}, the FGP is applied to a massive sea surface temperature dataset from NASA's Terra and Aqua satellites. Section \ref{sec: conclusion} is concluded with a brief summary and discussion on possible extensions.

\section{A Fused Gaussian Process Model} \label{sec: model}
This section starts with the definition of a fused Gaussian process (FGP) and then discusses its properties and relationship with other state-of-the-art methods. 
\subsection{Model Specification}
Suppose we are interested in a hidden real-valued spatial process $\{Y(\mathbf{s}): \mathbf{s} \in \mathcal{D}\subset \mathbb{R}^d\}$ in the spatial domain $\mathcal{D}$. Statistical inferences are  made upon the observed data $\bZ \equiv (Z(\s_1), \ldots, Z(\s_n))'$ with measurement error incorporated:
\begin{eqnarray} \label{eqn: measurement error}
Z(\s) =Y(\s) + \epsilon(\s),\, \s \in \mathcal{D},
\end{eqnarray}
where $\epsilon(\cdot)$ is a Gaussian white noise with mean zero and variance $\sigma^2_{\epsilon}v(\cdot)$. The quantity $v(\cdot)$ is known from validation data or instrument specification and allows the possibility of nonconstant measurement-error variances. The variance parameter $\sigma^2_{\epsilon}$ can also be inferred from validation data \citep[e.g.,][]{Cressie2006, Cressie2008, Nguyen2012}. If it is unknown, it will be estimated in practice. To model the hidden process $Y(\cdot)$, we assume the following structure,
\begin{eqnarray} \label{eqn: mean structure}
Y(\s) = \mu(\s) + \nu(\s),\, \s \in \mathcal{D},
\end{eqnarray}
where $\mu(\cdot)$ models the trend. In the remainder of this paper, we assume that $\mu(\cdot)=\X(\cdot)' \boldbeta$ with a vector of known covariates $\X(\cdot) = (X_1(\cdot), \ldots, X_p(\cdot))'$ and corresponding unknown coefficients $\boldbeta$. A zero-mean Gaussian process is assumed for the second term $\nu(\cdot)$  in (\ref{eqn: mean structure}) with a covariance function $C(\cdot,\cdot)$. Instead of specifying $C(\cdot,\cdot)$ directly, $\nu(\cdot)$ is assumed to be induced by two independent random vectors: a low-dimensional vector $\boldsymbol\eta\equiv (\eta_1, \ldots, \eta_r)'$ of size $r$ ($r\ll n$) and a high-dimensional vector $\boldsymbol \xi\equiv (\xi_1,\ldots, \xi_M)'$ of size $M$ ($M\approx n$ or $M> n$): 
\begin{equation}\label{eq:nu}
\nu(\s)=\sum_{i=1}^r S_i(\s)\eta_i + \sum_{j=1}^M A_j(\s) \xi_j \equiv \mathbf{S}(\mathbf{s})' \boldsymbol \eta
+ \mathbf{A}(\mathbf{s})' \boldsymbol \xi,
\end{equation}
where $\mathbf{S}(\cdot) \equiv (S_1(\cdot), \ldots, S_r(\cdot))'$ and $\mathbf{A}(\cdot) \equiv (A_1(\cdot), \ldots, A_M(\cdot))'$ are two sets of basis functions associated with $\bfeta$ and $\bfxi$, respectively. $\bS(\cdot)'\bfeta$ is called \textit{low-rank component}, and $\bA(\cdot)' \bfxi$ is called \textit{Gaussian-graphical-model component}. Their model specifications are presented below, respectively.

\noindent\textit{Low-rank Component $\bS(\cdot)' \bfeta$}

The $r$-dimensional vector $\bfeta$ is assumed to be a Gaussian random vector with zero mean and an unknown $r\times r$ covariance matrix $\bK$. The associated $r$ basis functions $\bS(\cdot)$ are fixed and \textit{known}. Such a low-rank component via a basis expansion has been widely used to analyze large spatial data sets. Various forms of basis functions have been suggested, including local bisquare functions \citep{Cressie2008}, wavelets \citep{Shi2007}, cubic B-splines \citep{Chu2014}, and basis functions resulted from a prespecified parametric covariance function and a set of prespecified locations or knots \citep{Banerjee2008}, among the others. In addition to prespecify basis functions, numerical methods for Karhunen-Lo\`eve (K-L) expansion have been developed to calculate eigenfunctions, which can be used as basis functions as suggested in \cite{Hu2013}. Recently, \cite{Bradley2016} point out that any class of basis functions can be re-weighted and then viewed as eigenfunctions within a K-L expansion, although sensitivity analysis is recommended to choose basis functions. Here, we choose multiresolutional and compactly supported basis functions for $\mathbf{S}(\cdot)$ as suggested in \cite{Cressie2008}. As demonstrated in previous works, incorporating a low-rank basis expansion enables dimension reduction, and thus is able to facilitate computationally feasible inference. Further discussion of the relationship and differences between our model and other models is given in the end of this section.
 
 \noindent\textit{Gaussian-graphical-model component $\bA(\cdot)'\bfxi$}

The $M$-dimensional vector $\bfxi$ is assumed to be a Gaussian random vector with zero mean and nonsingular covariance matrix $\bfSigma$. Suppose that $\bQ\equiv \bfSigma^{-1}=(q_{ij})$ denotes the corresponding precision matrix with $(i,j)$-th element $q_{ij}$. Then the Gaussian random vector $\bfxi$ can be represented by an undirected Gaussian graphical model via an undirected graph $\mathcal{G}=(\mathcal{V}, \mathcal{E})$, where $\mathcal{V}$ contains the $M$ vertices corresponding to the variables in $\bfxi$, and the edges $\mathcal{E}=(e_{ij})_{1\leq i < j\leq M}$ indicate whether the variables $\xi_i$ and $\xi_j$ ($i\neq j$) are conditionally independent given all other variables in $\bfxi$. Therefore, the variables in $\bfxi$ are Markov with respect to $\mathcal{G}$:
$$p(\xi_i|\{\xi_j:j\neq i\} ) = p (\xi_i| \{\xi_j: j\in \mathcal{N}_i\}),$$
where $p(\cdot)$ represents the probability density function; $\mathcal{N}_i\equiv \{ j| j\in \mathcal{V}, \text{ and }\{i,j\}\in \mathcal{E}\}$, and elements in the precision matrix $\bQ$ are non-zero only for neighbors and diagonal elements: $q_{ij}=0\iff  j\notin \mathcal{N}_i$ and $i\neq j$. In spatial statistics, an uGGM is also called Gaussian Markov random field \citep[e.g.,][]{Rue2005}. Note that it is also possible to define a directed GGM (dGGM) for $\bfxi$. However, since a dGGM can be converted to an uGGM via \textit{moralization} (i.e., ``marriage'' of a child node's parent nodes) \citep[e.g.,][]{Jordan2003}, an uGGM for $\bfxi$ is assumed in all numerical examples in this article.

The precision matrix $\bQ$ plays an important role in determining the dependence structure in $\bfxi$ and is usually modeled as a large structured sparse matrix, known up to a few parameters. For example, an explicit parametric form for $\bQ$ is provided in \cite{Lindgren2011} when a Mat\'ern covariance function is assumed.  Another commonly used approach is to assume a conditional autoregressive (CAR) structure \citep[sec.~4.2]{Cressie2011}. The covariance matrix of $\bfxi$ in a CAR model takes the form:
$\bfSigma\equiv \tau^2 (\bI - \gamma \bH) ^{-1} \bfDelta$,
 or equivalently, $\bQ\equiv \tau^{-2}\bfDelta^{-1}(\bI - \gamma \bH)$. Here the parameter $\gamma$  is interpreted as the strength of spatial dependence, while $\tau^2>0$ is a scale parameter; $\bH\equiv (h_{ij})$ is a known $M\times M $ matrix with zero diagonal elements; $\bfDelta\equiv diag(\Delta_1, \ldots, \Delta_M)$ is a known $M\times M$ diagonal matrix with positive diagonal elements. Meanwhile, to ensure that $\bQ=\tau^{-2}\bfDelta^{-1}(\bI - \gamma \bH)$ is symmetric and positive-definite, the parameter $\gamma$ needs to be restricted between the reciprocal of smallest and largest eigenvalues of $\bH$ (e.g., \citet{Besag1974}). This CAR model on $\bfxi$ implies the following conditional distributions,
 \begin{eqnarray} \label{eqn: GMRF}
\xi_i | \boldxi_{-i} \sim \mathcal{N}\left(\gamma\sum_{j=1}^M h_{ij} \xi_j,\, \tau^2\Delta_{i}\right), i=1,\ldots, M,
\end{eqnarray}
where $\boldsymbol \xi_{-i} \equiv (\xi_1, \ldots, \xi_{i-1}, \xi_{i+1}, \ldots, \xi_M)'$. Notice that when $\gamma=0$, $\bfxi\sim \mathcal{N}_M(\bfzero, \tau^2\bfDelta)$, which results in independence for $\{\xi_i: i=1,\ldots, M\}$.

Suppose that the random vector $\bfxi$ is defined on a generic lattice over the domain of interest, $\mathcal{D}\equiv \cup \{\mathcal{R}_i: i=1,\ldots, M\}$, where the $M$ small areal regions $\{\mathcal{R}_i\}$ are nonoverlapping, and called basic areal units (BAUs). To specify the basis functions $\bA(\cdot)\equiv (A_1(\cdot), \ldots, A_M(\cdot))'$, we define $A_i(\mathcal{R}_j)$ to be 1 if $i=j$ and zero otherwise. This discretization procedure is typically determined by the resolution of data so that each observation location is contained in one small areal region. The matrices $\bH$ and $\bfDelta$ are also specified according to the neighborhood structure in the lattice. In the following numerical examples, $\bH$ is constructed from first order neighborhood structure, and $\bfDelta$ is chosen to be the identity matrix.  In practice, $\{\mathcal{R}_i\}$ can be determined by the finest resolution for which spatial predictions will be made, and choice of neighborhood structure can be made based on model selection criteria such as cross-validation or Bayesian information criterion. We assume that the observed data are obtained at the same or a coarser spatial resolution compared to the lattice $\{\mathcal{R}_i: i=1,\ldots, M\}$, and thus $M$ can be much larger than the size of the data $n$. Such a lattice structure has been introduced and utilized to analyze remote-sensing data in previous studies \citep[e.g,][]{Nguyen2012}. 

The resulting model for the hidden process $Y(\s)$ is given by:
\begin{eqnarray}\label{eq:fgp}
Y(\s) =\bX(\s)'\bfbeta +\bS(\s)'\bfeta +\bA(\s)'\bfxi, 
\end{eqnarray}
which is called the fused Gaussian process (FGP), since it combines the low-rank and graphical-model components. Notice that the covariance matrix obtained from $Y(\cdot)$ at a finite collection of spatial locations has a fixed rank no more than $M$. The FGP methodology can also be viewed as a fixed rank kriging methodology.

\subsection{Connection with Several Existing Methods}
We include in this section remarks to compare the FGP with several models in literature for analyzing large or massive spatial data. The model in \eqref{eq:fgp} contains a low-rank component similar to that in \cite{Cressie2008}. As pointed out in later work \citep[e.g.,][]{Stein2014}, the performance of low-rank methods can be sensitive to the number of basis functions.  In Section \ref{sec: numerical examples}, we compare the predictive performance of FGP with FRK, and show that by introducing the GGM component, FGP is able to provide more accurate spatial predictions even with a fewer number of basis functions in the low-rank component than FRK does. 

It is worth noting that three other recently suggested methods for massive spatial data, the nearest-neighbor Gaussian process (NNGP) model in \cite{Datta2015}, the multi-resoution approximation (MRA) model in \cite{Katzfuss2017}, and Lattice Krig in \cite{Nychka2015} can all be considered as models induced from a GGM component. In particular, NNGP is induced by a directed Gaussian graphical model with vertices related to locations in a pre-specified reference set, while MRA is built based on a multi-resolution Gaussian graphical model but has assumed independence between vertices from different resolutions and vertices from different clusters (or regions) within the same resolution. Similarly,  in Lattice Krig, when we treat all grid cells as vertices in a graph, the model for random effects in Lattice Krig can be viewed as a Gaussian graphical model. In FGP, it is also possible to include a GGM component as those in NNGP, MRA, or Lattice Krig, especially when a specific target parametric covariance function is desired. Therefore, NNGP, MRA and Lattice Krig can be viewed as alternative parameterization for the GGM component in FGP, and the inference procedure and related computational advantages we present in Section \ref{sec: inference} will still hold. Many methods have been proposed recently to tackle the big ``$n$'' problem in spatial statistics \citep[e.g.,][]{Sang2012, Nychka2015, Datta2015, Katzfuss2017}. The FGP model is another competing method to tackle this computational problem, and at the same time FGP can model nonstationary spatial processes.

\section{Inference} \label{sec: inference}
This section details the inference procedure for parameter estimation and spatial predictions with FGP (Section \ref{sec: estimation and prediction}) and discusses the associated computational complexity (Section \ref{sec: computation}).

\subsection{Parameter Estimation and Spatial Prediction} \label{sec: estimation and prediction}

Let $\boldtheta$ denote the vector consisting of parameters in $\{\boldbeta, \K, \tau^2, \gamma\}$. Recall that the observed data is $\Z \equiv (Z(\s_1), \ldots, Z(\s_n))'$. By combining (\ref{eqn: measurement error}) and (\ref{eq:fgp}) and assembling vectors into matrices, the spatial linear mixed effects model for $\Z$ can be written as the following matrix form:
$$\Z=\X\bfbeta+\S\bfeta+\A\bfxi + \bfepsilon,$$
where  $\X\equiv [\bX(\s_1), \ldots, \X(\s_n)]'$ is the $n\times p$ matrix corresponding to the fixed effects $\bfbeta$; $\S\equiv[\S(\s_1), \ldots, \S(\s_n)]'$ is the  $n\times r$ matrix related to $r$-dimensional random vector $\bfeta$ in the low-rank component; $\A\equiv[\A(\s_1), \ldots, \A(\s_n)]'$ is the $n\times M$ matrix related to the $M$-dimensional vector $\bfxi$ in the uGGM component. Up to an additive constant, the negative log-likelihood function is written as:
$$l_Z(\bftheta)=\frac{1}{2}\{(\Z-\X\bfbeta)'\bC^{-1}(\Z-\X\bfbeta) + \log |\bC|\} + \text{constant},$$
where $\bC\equiv \text{var}(\Z)=\S\bK\S' + \A\Q^{-1} \A'+  \V_{\epsilon}$ with $\V_{\epsilon}=diag(\sigma^2_\epsilon v(\s_1), \ldots, \sigma^2_\epsilon v(\s_n))$. Evaluation of the negative log-likelihood function requires calculation of the inverse and log-determinant of the $n\times n$  matrix $\bC$, and it can be obtained efficiently using results in Proposition~\ref{prop: woodbury}. 
\begin{proposition} \label{prop: woodbury}
 Recall that $\bC\equiv \text{var}(\Z)=\S\bK\S' + \A\Q^{-1} \A'+  \V_{\epsilon}$, it follows that:
\begin{eqnarray}\label{eqn:Cinverse}
\bC^{-1}&=&\bD - \bD \bS (\bK^{-1} + \bS'  \bD \bS)^{-1} \bS' \bD, \\
\label{eqn:Cdet}
\log|\bC|&=&\log|\K^{-1}+\S'\D\S| + \log|\K| + \log|\D^{-1}|,
\end{eqnarray}
where $\D \equiv (\A\Q^{-1}\A' + \V_{\epsilon})^{-1} 
	= \V_{\epsilon}^{-1} - \V_{\epsilon}^{-1} \A 
	 (\Q + \A' \V_{\epsilon}^{-1}  \A)^{-1} \A' \V_{\epsilon}^{-1}$, and $\log|\D^{-1}| = \log | \Q + \A' \V_{\epsilon}^{-1}  \A| 
- \log |\Q| + \log | \V_{\epsilon} |$.
\end{proposition}
Proof of Proposition~\ref{prop: woodbury} can be found in Appendix~E. Note that the right-hand sides of \eqref{eqn:Cinverse} and \eqref{eqn:Cdet} involve only inversion and determinant of $r\times r$ matrices and $M \times M$ sparse matrices, which enable fast evaluation of the negative log-likelihood. 

To minimize the negative log-likelihood function $l_Z(\bftheta)$, iterative algorithms are devised. For example, as suggested in \cite{Chu2014}, the matrix $\K$ is parametrized via its eigendecomposition and then a two-step iterative algorithm can be carried out. Their theoretical results show that the parameters can be estimated consistently under certain regularity conditions.  First, $l_Z(\bftheta)$ is minimized with respect to the eigenvectors of $\bK$ for a fixed $(\bfbeta, \tau^2, \gamma)$ using a Newton-Raphson-type algorithm on a Stiefel manifold \citep{Peng2009}. Second, with fixed eigenvectors of $\bK$, $l_Z(\bftheta)$ is minimized with respect to the remaining parameters. Another technique to minimize $l_Z(\bftheta)$ is to treat random effects as ``missing data'', and the expectation-maximization (EM) algorithm \citep{Dempster1977} can be implemented. To devise the EM algorithm for the FGP, the random vector $\bfeta$ is treated as ``missing data''. Let $\bftheta_t$ denote the parameters at the $t$-th iteration. In the expectation step (E-step), the conditional expectations and covariance matrix for $\bfeta$ given the data $\Z$ and parameter estimates $\boldtheta_t$ are derived, respectively:
\begin{eqnarray*}
\boldmu_{\boldeta | \Z, \boldtheta_t} = E(\boldeta| \Z, \boldtheta_t) = \K_t \S' \bC_t^{-1} (\Z-\X\bfbeta_t) \text{ and }
\boldSigma_{\boldeta| \Z, \boldtheta_t} = \text{Var}(\boldeta| \Z, \boldtheta_t) = \K_t - \K_t \S' \bC_t^{-1} \S \K_t',
\end{eqnarray*}
where $\bC_t\equiv \S\K_t\S' + \A\Q^{-1}_t\A' + \V_\epsilon$, and $\Q_t \equiv \boldDelta^{-1}(\mathbf{I} - \gamma_t \H)/\tau^2_t$. In the maximization step (M-step), $\bftheta_{t+1}$ is updated by maximizing the so-called $Q$ function obtained in the E-step. In particular, closed-form updates can be derived for $\bK$ and $\bfbeta$, and numerical optimization procedures such as interior point method and active-set method \citep[e.g.,][]{Byrd1999}, are implemented to update $\tau^2$ and $\gamma$. 
Complete derivation of the EM algorithm for the FGP is included in Appendix~F as well as recommendations about initial values and convergence criteria. 

For spatial prediction, suppose that we are interested in making prediction of $Y(\cdot)$ at a set of locations $\{ \s_i^P\}_{i=1}^m\subset \mathcal{D}$ based on observed data $\Z$. Let $\Y^P \equiv (Y(\s_1^P), \ldots, Y(\s_m^P))'$. Define $\X^P \equiv [\X(\s_1^P), \ldots, \X(\s_m^P)]'$,  $\S^P \equiv [\S(\s_1^P), \ldots, \S(\s_m^P)]'$, and $\A^P=[\A(\s_1^P), \ldots, \A(\s_m^P)]'$. Conditioning on the parameter vector $\bftheta$, the predictive distribution, $\Y^P|\Z$, is derived as:  
\begin{eqnarray} \label{eqn: predictive dist}
\Y^P |\Z \sim \mathcal{N}_m( \X^P \boldbeta + \S^P \bfmu_{\bfeta|\Z} + \A^P \bfmu_{\bfxi|\Z},\, \bfSigma_{\Y^P|\Z}),
\end{eqnarray}
where \vspace{-0.5cm}
\begin{eqnarray*}
\bfmu_{\bfeta|\Z} &\equiv& \K\S'\bC^{-1}(\Z-\X\boldbeta), \\
\bfmu_{\bfxi|\Z} &\equiv& \Q^{-1}\A'\bC^{-1}(\Z-\X\boldbeta),  \\
\bfSigma_{\Y^P|\Z} &\equiv&  \S^P \boldSigma_{\boldeta|\Z}\S^{P'} 
+ \A^P\boldSigma_{\boldxi|\Z}\A^{P'} + \S^P\boldSigma_{\boldeta, \boldxi|\Z} \A^{P'} + (\S^P\boldSigma_{\boldeta, \boldxi|\Z} \A^{P'})',
\end{eqnarray*}
with $\boldSigma_{\boldeta|\Z}\equiv\text{var}(\boldeta|\Z) = \K- \K\S'\bC^{-1}\S \K'$, $\boldSigma_{\boldxi|\Z}\equiv\text{var}(\boldxi|\Z) = \Q^{-1} - \Q^{-1} \A'\bC^{-1}\A\Q^{-1}$, and $ \boldSigma_{\boldeta, \boldxi|\Z}\equiv \text{cov}(\boldeta, \boldxi| \Z)=-\K\S'\bC^{-1}\A\Q^{-1}$. Notice that the mean function in the posterior predictive distribution of $\Y^P |\Z$ gives the (simple) kriging predictor and the diagonal elements of the covariance matrix in the posterior predictive distribution of $\Y^P |\Z$ gives the kriging standard errors. In subsequent discussion, the kriging standard errors are simply referred to standard errors obtained from a predictive distribution.

\subsection{Computational Complexity} \label{sec: computation}
The main computational effort for inferences described in Section \ref{sec: estimation and prediction} is devoted to calculating the inverse and log-determinant of the $n\times n$ matrix, $\bC\equiv \S\bK\S' + \A\Q^{-1} \A'+  \V_{\epsilon}$, in which $\bK$ is only an $r\times r$ matrix with $r\ll n$, and $\Q$ is an $M\times M$ sparse matrix with $M\approx n$ or $M>n$. Using the results in Proposition~\ref{prop: woodbury}, such calculation solely involves inversion and determinant of $r\times r$ matrix and $M\times M$ sparse matrix. The former has computational complexity $O(r^3)$, while for the latter, the computation can be further reduced to calculate the Cholesky factor of the sparse matrix. The Cholesky factorization of a generic $M\times M$ matrix requires computational cost $O(M^3/3)$ and memory cost $O(M^2)$. As noted in \cite{Rue2005}, to calculate Cholesky factor of an $M\times M$ sparse matrix defined through an undirected GGM, efficient algorithms can be utilized to reduce the computational complexity to $O(M^{1.5})$ in two dimensional space, and its Cholesky factor requires memory cost $O(M\log M)$. As we focus on the CAR structure in the GGM component, the matrix $\A'\V_\epsilon^{-1} \A$ will be diagonal, and hence the matrix $\Q+\A'\V_\epsilon^{-1} \A$ has the same sparsity pattern and computational cost as the matrix $\Q$. For a single prediction location, the calculating the conditional mean and variance requires the computations of $\Q^{-1}\mathbf{a}, (\Q+\A'\V_\epsilon^{-1} \A)^{-1} \mathbf{T}, \mathbf{T}'(\Q+\A'\V_\epsilon^{-1} \A)^{-1} \mathbf{T}, \S'\D\S$ for a vector $\mathbf{a}$ of length $M$ and an $M\times r$ matrix $\mathbf{T}$, which require computational cost $O(M^{1.5})$, $O(M^{1.5}+M^{1.5}r)$, $O(M^{1.5}+M^{1.5}r + Mr^2)$, and $O(M^{1.5}+M^{1.5}r + Mr^2+nr^2)$, respectively. 
The overall computational cost is $O(M^{1.5}r+Mr^2)$. In terms of memory, inference for parameter estimation and spatial prediction requires to store the matrices $\bS$, $\bA$, Cholesky factors for $\Q$ and $\Q+\A'\V_\epsilon^{-1} \A$. In particular, the basis matrix $\S$ is sparse with memory cost less than $O(nr)$. The basis matrix $\A$ is also a sparse matrix with memory cost $O(n)$. So, these basis matrices have memory cost less than $O(nr)$. The memory cost for Cholesky factors of $\Q$ and $\Q+\A'\V_\epsilon^{-1} \A$ is also very cheap, since it only requires memory cost $O(M\log M)$. Although inversion of an $M\times M$ sparse matrix $ \Q+\A'\V_\epsilon^{-1} \A$ is needed, there is no need to store its inverse, but we only need to deal with much smaller matrices, $(\Q+\A'\V_\epsilon^{-1} \A)^{-1} \bT$, where $\bT$ represents an $M\times r$ matrix or $M$-dimensional vector. Thus, this matrix never has memory cost more than $O(Mr)$. Unlike the full Gaussian process, the overall memory cost in FGP will never exceed $O(M r)$ since $\log M$ is much smaller than $r$.

\section{Synthetic Examples} \label{sec: numerical examples}
In this section, several simulation examples are provided to demonstrate the robustness under misspecified covariance families and nonstationary performance of FGP as well as its computational efficiency.  We apply FGP to analyze a massive sea surface temperature (SST) dataset, and make comparisons with other existing methods as well. In all examples, the FGP  is implemented in MATLAB R2018b with the MATLAB function \emph{fmincon} used for numerical optimization. The simulation studies are carried out on a 4-core HP system with Intel Xeon x5650 CPU and 12 Gigabytes memory. The analysis of the SST data is carried out using 16 cores, and computation for the variance of predictions over entire region of interest is accomplished through parallel computing. Several figures are generated using the \texttt{ggplot2} package \citep{Wickham2016} and the \textsf{R} software \citep{R}.

To evaluate the predictive performance in the following simulation examples, the root-mean-squared-prediction error (RMSPE) is used to compare out-of-sample predictions. To quantify the validility of predictive distributions, the continuous-rank-probability score \citep[CRPS;][]{Gneiting2007} is reported, where small values of CRPS indicate better prediction results.

\subsection{Robustness under stationary covariance families} \label{subsec: robust examples}

The goal of this section is to demonstrate the robustness of predictive performance for FGP when the underlying true processes are simulated with different families of covariance functions. In particular, we only focus on Mat\'ern covariance functions and a spherical covariance function with the following forms:
\begin{align} \label{eqn: covariance function}
\begin{split}
    \text{Mat\'ern: } & C(h) = \sigma^2 \frac{2^{1-\nu}}{\Gamma{(\nu)}}\left( \sqrt{2\nu} \frac{h}{\rho} \right) \mathcal{K}_{\nu}\left(\sqrt{2\nu}\frac{h}{\rho}\right), \\
    \text{Spherical: }& C(h) = \sigma^2\left\{ 1 - 1.5\left(\frac{h}{\rho}\right) + 0.5\left(\frac{h}{\rho}\right)^3 \right\} I (h<\rho),
    \end{split}
\end{align}
where $\Gamma(\cdot)$ is the gamma function, $\mathcal{K}_{\nu}(\cdot)$ is the modified Bessel function of the second kind with order $\nu$. $\sigma^2$ is called partial sill, $\rho$ is the range parameter, and $\nu$ is the smoothness parameter in the Mat\'ern covariance function. $I(h<\rho)$ is one if $h< \rho$ and zero otherwise.

As a benchmark, two alternative methods are implemented in addition to the FGP model. The first alternative is to perform kriging using the true covariance function, which is referred to as \textit{EK}. The second alternative is to perform kriging with an exponential covariance function regardless of the true underlying covariance structure, which is referred to as \textit{MK}. Here, we only consider the situation that other types of covariance functions are misspecified as exponential covariance function, i.e., Mat\'ern with $\nu=0.5$. In addition, we also compare FGP with Lattice Krig and FRK, which are implemented based on the \textsf{R} packages \texttt{LatticeKrig} \citep{Nychka2016} and \texttt{FRK} \citep{Zammit2017}.

We simulate $M=2500$ data points in the domain $\mathcal{D}\equiv [0, 50]\times [0, 50]$. Two scenarios are considered. In each scenario, three different covariance functions are used to simulate the underlying true spatial processes: Mat\'ern with $\nu=0.5$, Mat\'ern with $\nu=2$, and spherical covariance function. In Scenario 1, we fix the effective range parameter to be 20 in the Mat\'ern covariance function for $\nu=0.5$ and $\nu=2$. The range parameter in the spherical covariance function is chosen to be $10$. The partial sill is set to be 16 and the measurement-error variance is set to be 1.6, i.e., signal-to-noise ratio (SNR) is 10. In Scenario 2, we fix the effective range parameter to be 45 in the Mat\'ern covariance function for $\nu=0.5$ and $\nu=2$. The range parameter in the spherical covariance function is chosen to be 20. The partial sill is also fixed at 16 with SNR at 10.

In each scenario, 10\% randomly selected data are held out to assess the short-range predictive performance. We compared EK, MK, FGP, FRK, and Lattice Krig based on 30 repeated simulation runs, where the measurement error process is generated with 30 different random seeds. To implement EK, the \textit{true} covariance  parameters $\phi$, $\sigma^2$, and $\sigma^2_\epsilon$ are used to perform kriging with the full covariance structure, which should give the best predictions and can be used as a baseline in each simulation experiment. To implement MK, an exponential covariance function model is assumed, and the parameters are estimated via maximum likelihood methods based on observed data, and predictions are obtained through kriging. The FRK is implemented based on the \textsf{R} package \texttt{FRK} with two resolutions of basis functions, since adding additional basis functions in FRK gives much worse results with the \texttt{FRK} package. So, the low-rank component in FGP is fitted with bisquare basis functions at two different resolutions obtained from the \textsf{R} package \texttt{FRK}. The GGM component in FGP is assumed to follow a CAR model with its proximity matrix $\H$ constructed with the first order neighborhood structure. The EM algorithm described in Section \ref{sec: estimation and prediction} is utilized to estimate parameters and obtain spatial predictions based on formulas \eqref{eqn: predictive dist}. Lattice Krig is implemented based on the \textsf{R} package \texttt{LatticeKrig} with three levels of basis functions and default settings for other tuning parameters.

\begin{figure}[htbp]
\renewcommand{\figurename}{Fig.}
\captionsetup{labelsep=period}
   \centering
  \makebox[\textwidth][c]{ \includegraphics[width=1.0\textwidth, height=0.4\textheight]{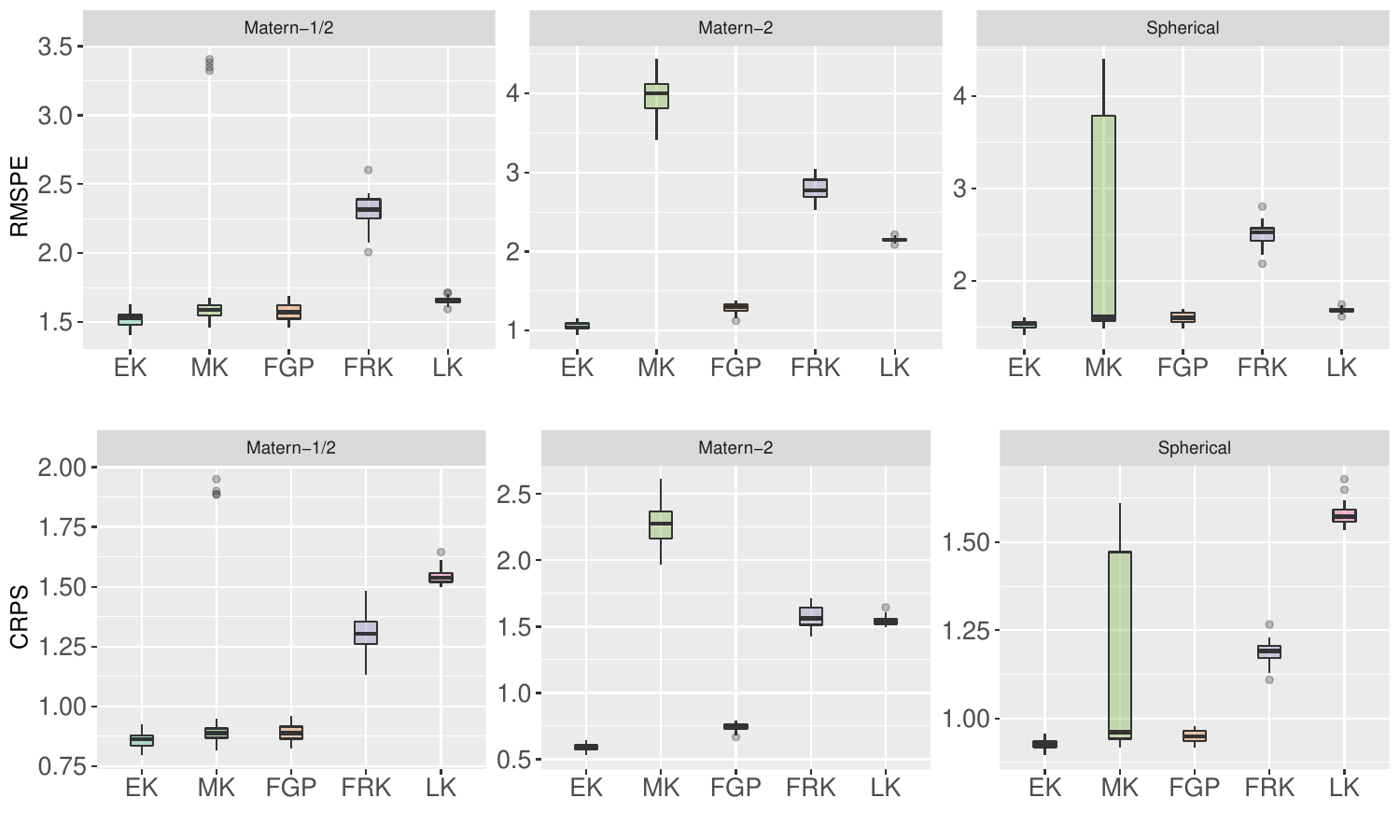}}%
   \caption{Boxplot of predictive measures in Scenario 1 based on 30 simulations. The top and bottom rows show the RMSPEs and CRPSs for five methods under three different covariance functions, respectively. Note that LK is short for Lattice Krig.}
   \label{fig: 2D scenario 1}
\end{figure}

The procedure EK with the true covariance model and true covariance parameters, performs the best among all methods. The prediction results in Scenario 1 are shown in Figure~\ref{fig: 2D scenario 1}. For effective range 20 in the Mat\'ern covariance function, the predictive performance of EK and FGP improves as the spatial process becomes smoother, while the predictive performance of MK and Lattice Krig deteriorates significantly. MK shows unstable predictive results under different covariance functions in terms of RMSPE and CRPS, since the performance of MK deviates significantly relative to EK under the Mat\'ern with $\nu=2$ and the spherical covariance function. FGP gives similar RMSPE and CRPS as those in EK. Lattice Krig gives a slightly larger RMSPE and much larger CRPS than FGP. FRK gives worst prediction results among all methods under all three different covariance functions. 

\begin{figure}[htbp]
\renewcommand{\figurename}{Fig.}
\captionsetup{labelsep=period}
   \centering
  \makebox[\textwidth][c]{ \includegraphics[width=1.0\textwidth, height=0.4\textheight]{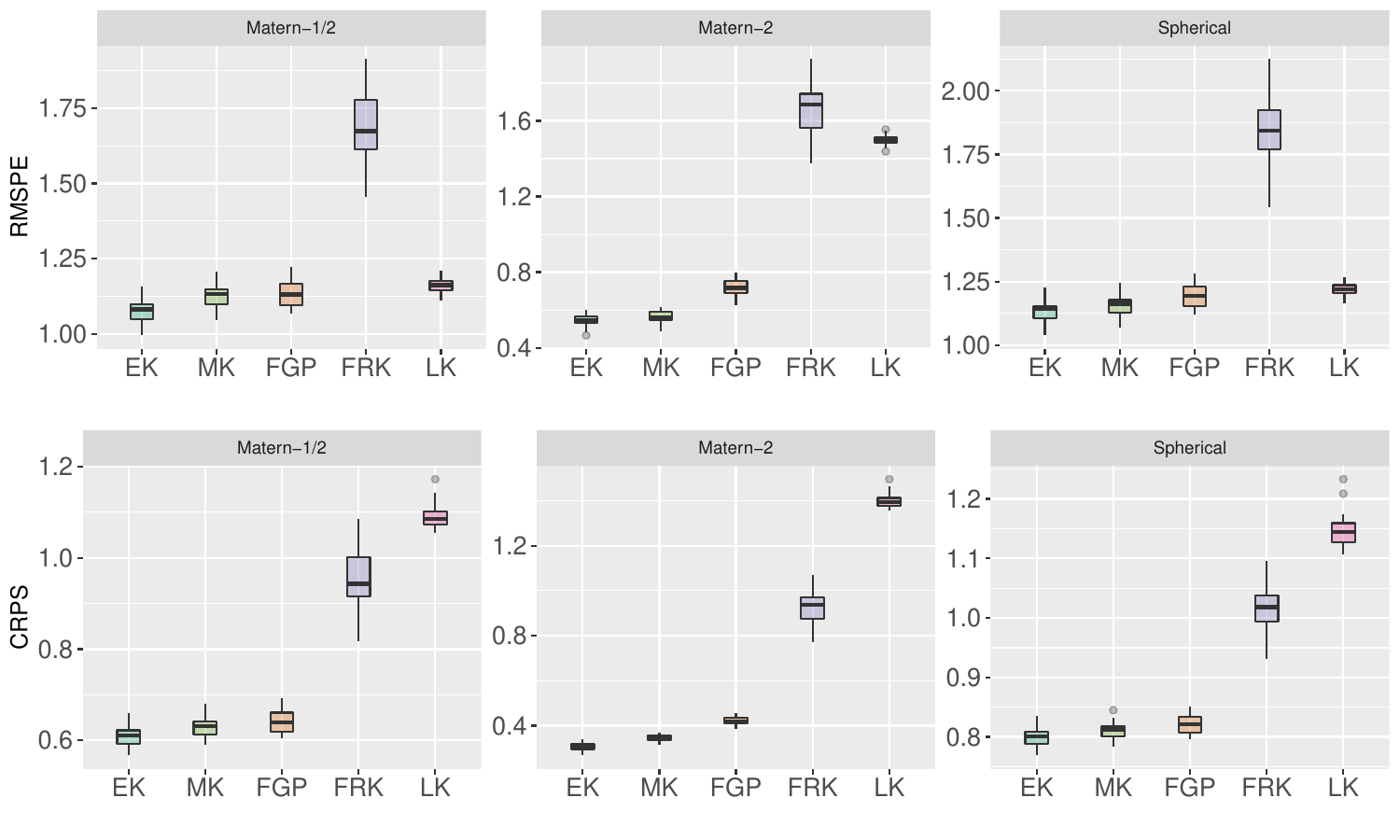}}%
   \caption{Boxplot of predictive measures in Scenario 2 based on 30 simulations. The top and bottom rows show the RMSPEs and CRPSs for five methods under three different covariance functions, respectively. Note that LK is short for Lattice Krig.}
   \label{fig: 2D scenario 2}
\end{figure}

When the effective range is increased to 45 in the Mat\'ern covariance function, Figure~\ref{fig: 2D scenario 2} show that EK, MK, and FGP give better results as the spatial process becomes smoother compared to Scenario 1, while Lattice Krig gives worse predictive results. MK performs much stable relative to EK than those in Scenario 1. It is not surprising that MK can perform well in practice \citep{Stein1988, Kaufman2013}. In contrast, FGP is also able to give similar predictive performance as MK and at the same time FGP gives better predictive performance than Lattice Krig. Lattice Krig gives larger CRPS than all the other methods. This is likely due to the fact that conditional simulation technique is used to approximate the predictive variance. Figure~\ref{fig: 2D scenario 2} shows that both MK and FGP achieve above 94\% median relative efficiency over EK under the Mat\'ern covariance with $\nu=0.5$ and the spherical covariance function. In contrast, Lattice Krig has slightly smaller median relative efficiency that FGP under these two covariance functions. For the Mat\'ern covariance with $\nu=2$, FGP does not have good performance as MK but does have better performance than Lattice Krig. The predictive performance of FRK seems to vary slightly when the smoothness parameter changes from 0.5 to 2. This probably explains why there is a moderate discrepancy between the predictive performance of FGP and that of EK.

In our experiments, we did not tune the basis functions in the low-rank component as well as those in the GGM component. With a fairly easy specification, FGP can give robust prediction results when the underlying true fields are generated by different stationary covariance functions.  In Appendix~B, we also show covariance approximations by evaluating correlation matrices over 50-by-50 grid points in these two scenarios. Both MK and Lattice Krig have similar patterns in their covariance matrix plots, {but FRK and FGP have different patterns. This is because the low-rank matrices in FRK and FGP are unstructured, and they don't assume a specific parametric model for the random coefficients as in Lattice Krig.} When the quality of likelihood approximation including covariance approximation is desired, one can impose a further parametric structure on the matrix $\mathbf{K}$. It is worth noting that the FGP model is not designed to approximate a target covariance function, but is designed to model the data directly regardless of the underlying true covariance structures and is aimed at predictive inference. Additional numerical results related to computing time and covariance approximation can be found in Appendix~A and Appendix~C.

\subsection{Nonstationary Performance} \label{subsec: nonstationary example}

This section aims at demonstrating the predictive performance of FGP under nonstationary spatial processes.

\begin{figure}[htbp]
\renewcommand{\figurename}{Fig.}
\captionsetup{labelsep=period}
\begin{center}
\makebox[\textwidth][c]{ \includegraphics[width=1.0\textwidth, height=0.35\textheight]{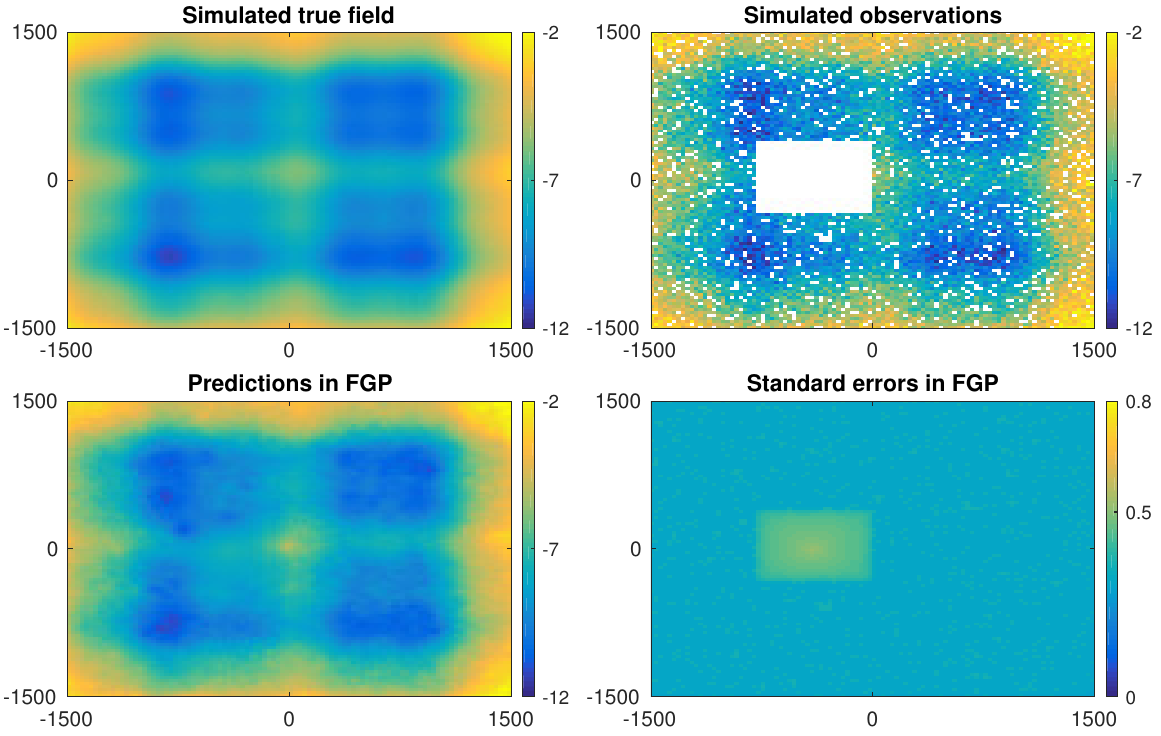}}
\caption{A simulated dataset and prediction results from FGP. The upper-left panel shows the underlying true field $Y(\cdot)$ evaluated at $100\times 100$ locations. The upper-right panel shows the observations by adding random measurement errors to  $Y(\cdot)$. Locations with observations held out are colored white. The bottom-left panel shows the spatial predictions from $Y(\cdot)$ in FGP, while the bottom-right panel plots the corresponding standard errors. \vspace{-1cm}}
\label{fig: 2d simulation example}
\end{center}
\end{figure}

To setup the experiment, we choose the following deterministic function $f(s) = \exp\{ - (s/750-1)^2 \} + \exp\{-0.8(s/750+1)^2\} - 0.05\sin(8(s/750+0.1))$ adapted from \cite{Guhaniyogi2017}. The hidden process $Y(\cdot)$ is generated from $f(\cdot)$ by defining $Y(\mathbf{s}) = -10 f(s_1) f(s_2)$ in the domain $\mathcal{D}=[-1500, 1500]\times [-1500, 1500]$, where $\mathbf{s}\equiv(s_1, s_2) \in \mathcal{D}$. We first generate the true process $Y(\cdot)$ at a $100\times 100$ regular grid in $\mathcal{D}$; see the upper-left panel of Figure~\ref{fig: 2d simulation example}. The data are then generated by adding a measurement-error process $\epsilon(\cdot)$ such that the signal-to-noise ratio is 10. Specifically, the empirical variance of $Y(\cdot)$ at these 10000 locations is calculated first, $\hat{\sigma}^2_Y=4.4343$. The measurement-error variance is then set to be $\sigma^2_{\epsilon}=0.1\hat{\sigma}^2_{Y}$. To evaluate prediction performance, we hold out data in the rectangular region $[-750, 0,]\times [-375, 375]$ to test long-range prediction skills. In addition, we hold out data at 1,000 randomly selected remaining locations to evaluate short-range prediction skills. The right-panel of Figure~\ref{fig: 2d simulation example} shows the remaining 8400 observations, which are used to fit the following six models: 
\begin{itemize}[noitemsep, topsep=0pt]
\item[(1)] CAR model with the first order neighborhood structure;
\item[(2)] NNGP with 25 neighbors and 15000 MCMC samples;
\item[(3)] Lattice Krig with $900+2401+7569=10870$ basis functions at three resolutions;
\item[(4)] FRK model with $r=24+136=160$ basis functions at two different levels;
\item[(5)] FRK model with $r=24+136+688=848$ three levels of basis functions at three different levels;
\item[(6)] FGP  with two levels of basis functions ($r=160$) and first order neighborhood structure in GGM component.
\end{itemize}
The CAR model and FGP  are implemented in MATLAB. NNGP is implemented with the \textsf{R} package \texttt{spNNGP} \citep{Datta2015} using the response option. Lattice Krig is implemented with \texttt{LatticeKrig} using the default setting. FRK is implemented with \texttt{FRK}. 

To compare each model, RMSPEs and CRPSs are computed based on 15 simulation runs. Table~\ref{table: 2D deterministic function} shows that FGP gives smallest RMSPE and CRPS among all the methods. In particular, the CAR model as a special implementation of FGP without the low-rank component gives the largest RMSPE. The NNGP model assumes an exponential covariance function, and gives the second largest RMSPE. These two models fail to capture the nonstationary spatial dependence structure. FRK is first fitted with two levels of basis functions. Its prediction accuracy is much better than that of CAR and NNGP. There are two reasons for NNGP model to give poor performance not as good as some other methods: the NNGP model approximates a stationary exponential covariance function which is deemed not to capture nonstationary dependence structure very well; and most prediction locations in the contiguous missing regions share the same neighbor sets due to the way NNGP works, which leads to a large discrepancy compared to the underlying true field.  For FRK, the predictive performance does not improve but becomes worse when a third resolution of basis functions is added. This indicates that adding more basis functions may not necessarily improve the predictive performance of FRK if the basis functions are not chosen well. This poor predictive performance in FRK is possibly due to some artifacts on the choice of basis functions over large contiguous missing regions as pointed out in \cite{Bradley2018}. Readers are referred to \cite{Tzeng2017} and \cite{Ma2017downscaling} for alternative ways to specify basis functions that improve the predictive performance in FRK. FGP performs better than FRK even though more basis functions are incorporated in FRK. Lattice Krig performs slightly better than FRK, but does not perform as well as FGP. We also report the computing time for these methods, though we do not recommend interpreting it literately for methods implemented in different softwares. The predictions and associated kriging standard errors from FGP are shown in Figure~\ref{fig: 2d simulation example}, which reveals that FGP can well capture the nonstationary structure for this smooth underlying true field. In Appendix~D, we also illustrate that FGP can also perform very well under a non-smooth underlying true field.   
\begin{table}[htbp]
\centering
\normalsize
   \caption{Results under a nonstationary spatial field. The average of RMSPEs and the average of CRPSs over $15$ simulations are reported for each model with standard deviations included in the parenthesis. The average computing time over 15 simulations for each model is also reported.}
  {\resizebox{1.0\textwidth}{!}{%
  \setlength{\tabcolsep}{1.8em}
   \begin{tabular}{l c c c c c c} 
   \toprule \noalign{\vskip 1.5pt}
\multirow{2}{*}{Model	} & \multirow{2}{*}{CAR}  & \multirow{2}{*}{NNGP}  & \multirow{2}{*}{Lattice Krig} & \multicolumn{2}{c}{FRK}  & FGP  \\ \cline{5-6} \noalign{\vskip 2pt}   \noalign{\vskip 1.5pt} 
	&  &    &    &   $r=160$ & $r=848$    & $r=160$ \\  
\midrule
\multirow{2}{*}{RMSPE}& 2.3492 & 1.5194	& 0.7331  & 0.6704 & 1.5023 & 0.4733 \\ \noalign{\vskip 1.5pt}  
				& (0.022)	& (0.076) 	& (0.019)	&  (0.053)	& (0.082) & (0.1059) \\ \noalign{\vskip 1.5pt}   \noalign{\vskip 1.5pt}
\multirow{2}{*}{CRPS}&1.2575   &1.3023	&0.7710	& 0.9721	& 1.1512 &0.3029 \\ \noalign{\vskip 1.5pt}  
				& (0.012)	& (0.141)	& (0.033)	& (0.081)	& (0.040) &(0.041) \\ \noalign{\vskip 1.5pt} \hline \noalign{\vskip 3pt} 
		Time (mins)  & 0.32 &63.6    &3.20     &0.31        &1.00         &3.50 \\		
 \bottomrule
   \end{tabular}%
   }}
   \label{table: 2D deterministic function}
\end{table}  


\section{Application with Sea Surface Temperature Data} \label{sec: application}

Sea surface temperature plays a vital role in the Earth's atmosphere and climate systems. A complete and accurate map of SST  is essential in oceanographic sciences, weather forecasts, and in studying global and regional climate changes \citep{Donlon2002}. In this section, the performance of FGP is illustrated by analyzing a dataset of $n=391,789$ observations of SST on July 6, 2002. These data are obtained by combining and transforming original Level 2 data at 4 or 9 km spatial resolutions from the Moderate Resolution Imaging Spectroradiometer (MODIS) instruments on board NASA's Terra and Aqua satellites. The resulting data product is defined at a global grid of equal-areal hexagon cells with intercell distance 30 km, called Discrete Global Grids \citep[DGGs;][]{Sahr2003}. This overlaid grid of 30-km hexagons defines the spatial resolution of interest in our analysis and is used for the corresponding lattice for the GGM component in FGP. There are 440,894 DGG hexagons over ocean, due to alignment of the satellite orbits and failure to retrieve (e.g., presence of clouds). SST observations are available at $n=391,789$ hexagons, and thus there are 49,105 hexagons without any observations. The measurement-error variance is estimated as 0.0112 by fitting empirical semivariograms near the origin as in \cite{Kang2010}. 

Exploratory analysis suggests that the trend term be modeled as a quadratic function in terms of latitude. The covariates $X(\cdot) = [1, \text{latitude}(\cdot),$ $\text{latitude}^2(\cdot)]'$ are used in \eqref{eq:fgp}. For the low-rank component, as suggested in \cite{Cressie2008}, the multi-resolutional local bisquare basis functions are used over the globe. These basis functions are defined as:
$S(\mathbf{u}) = \{1 - (\|\mathbf{u}-\mathbf{v}\|/r)^2\}^2$ if $\|\mathbf{u}-\mathbf{v}\|\leq r$, and $S(\mathbf{u})=0$ otherwise, where $\mathbf{v}$ is the center of basis function and $r$ is the radius of basis function. There are 32 basis functions from the first resolution, 92 from the second and 272 from third. Due to the fact that there are 138 basis functions with majority of support over land instead of ocean, these basis functions are excluded, which results in a total number of $r=258=21+61+176$ basis functions in the low-rank component. For the GGM component, the CAR model is assumed, where the proximity matrix $\H$ is constructed with 0-1 weights based on first order neighborhood structure, and $\boldDelta$ is chosen to be the identity matrix. 
 
In the validation study, data are held out in two ways: (1) 22,204 observations are held out in a specific region $\mathcal{S}_1$ between latitudes $-60^{\circ}$ and $60^{\circ}$ and longitudes $-145^{\circ}$ and $-130^{\circ}$, referred to as missing by design; (2) $10\%$ of remaining observations are randomly sampled. The corresponding set of locations is denoted by $\mathcal{S}_2$, referred to as missing at random. Therefore, $59,162$ observations are held out in $\mathcal{S}_1\cup \mathcal{S}_2$ to evaluate the predictive performance, and $391,789-59,163=332,626$ observations are used for parameter estimation. In what follows, we only compare FGP with Lattice Krig, FRK, and CAR models. For Lattice Krig, we use the default setting in \texttt{LatticeKrig}  with $(8,568+30,276+113,442=) 152,286$ basis functions at three different resolutions. For FRK, we first consider 258 basis functions at three different levels, and then we add additional fine-resolutional basis functions so that FGP and FRK will have similar computing time or memory cost. Specifically, we add basis functions at next two finer resolutions with a total of $r=258+614=872$ basis functions and $r=872+158=1030$ basis functions, respectively, where large number of basis functions at much finer resolution have been deleted due to numerical stability problems. We did not further increase the number of basis functions in FRK, since it requires too much computer memory. FGP is only implemented with 258 basis functions at the first three resolutions. The EM algorithms are used for both FGP and FRK to obtain parameter estimates. For the CAR model, the maximum likelihood estimates are obtained through numerical optimization using the MATLAB function \emph{fmincon} with the interior-point algorithm. To compare predictive performance of Lattice Krig, FRK, FGP, and CAR, the RMSPE and CRPS are calculated for these methods. The results in Table~\ref{table:MSPE_CV_SST} show that FGP outperforms the other three methods in terms of predictive performance no matter whether data are missing in a large region or missing at randomly sampled locations. Even though we increase the number of basis functions in FRK, FGP can still perform better than FRK. The parameter estimation in Lattice Krig took about 2.3 hours. The EM algorithms in FRK took about 8.0 minutes, 40 minutes, and 60 minutes to get parameter estimates for 258, 872, 1030 basis functions in FRK, respectively. The numerical optimization algorithm to obtain maximum likelihood estimates in the CAR model took 6.4 minutes. The EM algorithm for FGP took about 2.4 hours to get parameter estimates. The computation of spatial predictions at all prediction locations took about 41.0 seconds in FRK with 258 basis functions, 49.8 seconds in FGP, and 4.7 seconds in the CAR model. Computing the kriging standard error at each location took about 11.4 seconds in FRK, 64.1 seconds in FGP, and 4.9 seconds in the CAR model. This example also shows that with a more complicated model, FGP does require more computing time compared to the other two methods, but it is capable of providing better and more reliable predictions. 
 
\begin{table}[htbp]
\centering
   \caption{Results from the validation study using the SST data. The RMSPE and CRPS are given for Lattice Krig, FRK, FGP, and CAR over all held-out locations.}
  {\resizebox{1.0\textwidth}{!}{%
  \setlength{\tabcolsep}{1.8em}
   \begin{tabular}{ l c c c  c c  c } 
   \toprule \noalign{\vskip 1.5pt} 
		& \multirow{2}*{Lattice Krig}  & \multicolumn{3}{c}{FRK} & FGP &  \multirow{2}*{CAR} \\ \cline{3-5} \noalign{\vskip 1.5pt}  
	 	& & $r=258$ & $r=872$ & $r=1030$ & $r=258$ & \\ 
	 \midrule \noalign{\vskip 1.5pt} 
RMSPE  &0.6421	&  1.1292 & 1.0210 &0.8965 & 0.5749 & 0.6676 \\ \noalign{\vskip 1.5pt}  \noalign{\vskip 3pt} 
CRPS &0.8024  &0.5957   &0.5381 & 0.4731 & 0.3013   &0.5830 \\
\noalign{\vskip 1.5pt} \bottomrule
   \end{tabular}%
   }}
   \label{table:MSPE_CV_SST}
\end{table}

\begin{figure}[htbp]
\renewcommand{\figurename}{Fig.}
\captionsetup{labelsep=period}
   \centering
  \makebox[\textwidth][c]{ \includegraphics[width=1.0\textwidth, height=0.4\textheight]{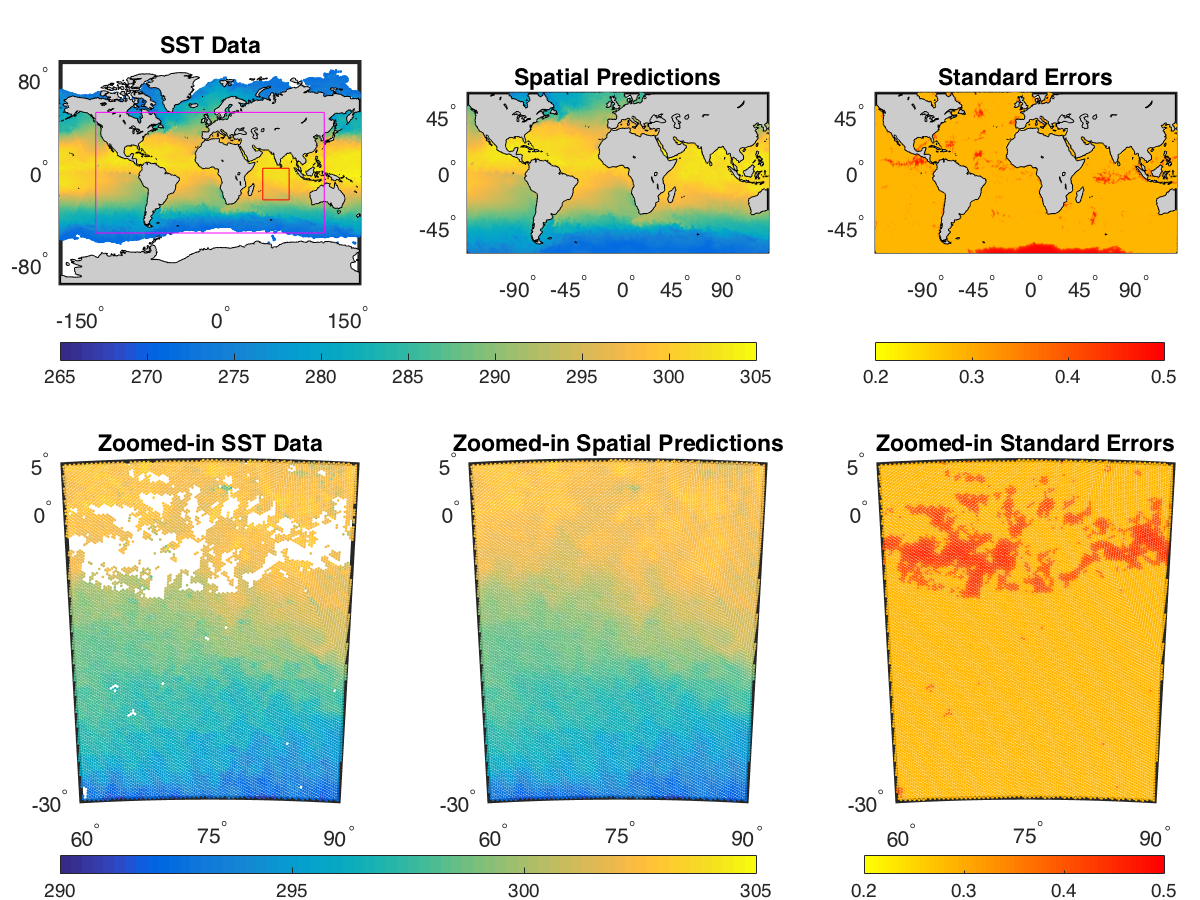}}%
   \caption{SST and its predictions. Upper-left panel: the SST data (unit: Kelvin). The larger rectangular region delineated shows the region where spatial predictions will be made. The smaller rectangular region delineated marks the area for which zoomed-in views are shown in the bottom panels. Upper-middle and upper-right panels: the spatial predictions and associated standard errors from the FGP, respectively.}
   \label{fig: FGP window}
\end{figure}

We now provide the results when FGP is used to analyze all available SST data. The estimates of conditional marginal variance and spatial dependence parameters in FGP are $\hat{\tau}^2=0.123$ and $\hat{\gamma}= 0.163$. Spatial predications are only made at 283,966 locations in a large rectangular region between longitudes $-130^{\circ}$ and $130^{\circ}$ and latitudes $-60^{\circ}$ and $60^{\circ}$ region over the entire ocean. The FGP took about 52.4 seconds to obtain all the spatial predictions over entire ocean. Figure~\ref{fig: FGP window} shows the predictions and associated kriging standard errors over this large region (upper panels) as well as zoomed-in maps for a subregion in Indian Ocean (lower panels), which clearly suggests that the standard errors reflect the pattern of the missing data, and that the FGP captures spatial variation of SST very well.

\section{Discussion} \label{sec: conclusion}

This article presents a fused Gaussian process model that combines the low-rank component and an undirected Gaussian graphical model to analyze very large spatial datasets. The covariance function in FGP allows nonstationary dependence structure and robust predictive performance. Numerical studies show that the FGP allows fast computation for very large or massive spatial datasets and is able to provide efficient and robust spatial predictions against misspecification of the spatial covariance structure. The current implementation of FGP seems to have limited ability to approximate a target covariance function, but this does not affect the validity and efficiency of its associated predictive distribution, since FGP is designed to model data directly with a semiparametric covariance function. If the quality of covariance approximations is the primary target, normalization of basis functions might help FGP to approximate stationary covariance functions as in \cite{Nychka2015}. In current examples, the graphical-model component is assumed with a parsimonious model: the spatial conditional autoregressive model. Alternative graphical models can be utilized for the GGM component in FGP. For example, the method in \cite{Lindgren2011} can be used to construct the precision matrix in the GGM component. The corresponding basis function can be chosen to be a piecewise linear basis function. The GGM component can also be chosen as a simultaneous autoregressive model, and the corresponding basis function can be chosen to be the Wendland basis function in a similar way as in \cite{Nychka2015}. 

A natural extension of a FGP model is to impose a special structure in the covariance matrix in the low-rank component, and perform fully Bayesian inference. For instance, one can assume a parametric covariance function such as the Mat\'ern family in the low-rank component. A fully Bayesian analysis can thus be carried out with appropriate prior specifications. Moreover, the GGM component in a FGP model can be assumed with independent block structures defined on partitioned subregions. This allows efficient computations for extremely large number of BAUs. These topics will be left for future research.

The FGP model has very nice change-of-support property. Let $\mathcal{R} \subset \mathbb{R}^d$ and define $Y(\mathcal{R})\equiv \int_{\mathcal{R}} Y(\s) \text{ d}\s/|\mathcal{R}|$, where $|\mathcal{R}|$ is the $d$-dimensional volume of $\mathcal{R}$. Then
\begin{eqnarray*}
\text{cov}(Y(\mathcal{R}_1), Y(\mathcal{R}_2)) = \S(\mathcal{R}_1)' \K \S(\mathcal{R}_2) + \A(\mathcal{R}_1)'\Q^{-1} \A(\mathcal{R}_2),\, \mathcal{R}_1, \mathcal{R}_2 \subset \mathbb{R}^d,
\end{eqnarray*}
where $\S(\mathcal{R}) \equiv (S_1(\mathcal{R}), \ldots, S_r(\mathcal{R}))'$; $\A(\mathcal{R})\equiv (A_1(\mathcal{R}), \ldots, A_M(\mathcal{R}))'$; $S_i(\mathcal{R}) \equiv \int_{\mathcal{R}} S_i(\s) \text{ d}\s/|\mathcal{R}|$; and $A_i(\mathcal{R}) \equiv \int_{\mathcal{R}} A_i(\s)\text{ d}\s/|\mathcal{R}|$ for $\mathcal{R}\subset \mathbb{R}^d$. Thus, the basis functions can be integrated offline and the formulas for spatial prediction and standard error will be of the same form. The current FGP model can be easily generalized to the space-time framework. For example, one may assume a spatio-temporal random-effect model $\nu(\s, t) = \S_t(\s)' \boldeta(t)$, where the basis function $\S_t(\cdot)$ depends on time $t=0, 1, \ldots$, and random vectors $\{\boldeta(t): t=0, 1, \ldots\}$ follow an autoregressive model \citep{Cressie2010}; see \cite{Ma2018DFGP} for detailed model formulation.

\section*{Supplementary Materials}
The online supplementary materials include additional numerical simulations and the details of the EM algorithm. In addition, we also include the computer code to implement the FGP model with illustrating examples in the simulation study.

\onehalfspacing
\setlength{\bibsep}{4pt}
\bibliographystyle{apa}
\bibliography{myspatial}{}

\begin{thebibliography}{}
\bibitem[\protect\astroncite{Bradley et~al.}{2018}]{Bradley2018}
Bradley, J.~R., Wikle, C.~K., and Holan, S.~H. (2018).
\newblock Hierarchical models for spatial data with errors that are correlated
  with the latent process.
\newblock {\em Statistica Sinica}.
\end{thebibliography}

\end{document}